\documentclass[
 nopreprint,
author-year,
]{jasatex}

\usepackage{graphicx}
\usepackage{dcolumn}
\usepackage{amsmath,amsfonts}
\usepackage{bm}

\begin{document}

\preprint{preprint}

\title[Reconstruction of Rayleigh-Lamb dispersion from noise]{Reconstruction of Rayleigh-Lamb dispersion spectrum based on noise obtained from an air-jet forcing }

\author{Eric LAROSE}


\author{Philippe ROUX}

\author{Michel CAMPILLO}

\affiliation{Lab. de G\'eophysique Interne et Tectonophysique, Universit\'e J. Fourier \& CNRS, BP53, 38041 Grenoble, France. Email: eric.larose@ujf-grenoble.fr}

\date{\today}

\begin{abstract}
The time-domain cross-correlation of incoherent and random noise recorded by a series of passive sensors contains the impulse response of the medium between these sensors. By using noise generated by a can of compressed air sprayed on the surface of a plexiglass plate, we are able to reconstruct not only the time of flight but the whole waveforms between the sensors. From the reconstruction of the direct $A_0$ and $S_0$ waves, we derive the dispersion curves of the flexural waves, thus estimating the mechanical properties of the material without a conventional electromechanical source. The dense array of receivers employed here allow a precise frequency-wavenumber study of flexural waves, along with a thorough evaluation of the rate of convergence of the correlation with respect to the record length, the frequency, and the distance between the receivers. The reconstruction of the actual amplitude and attenuation of the impulse response is also addressed in this paper.
\\
\\
  (accepted for publication in J. Acoust. Soc. Am. 2007)  
\end{abstract}

\pacs{4340Dx, 4335Cg, 4350Yw, 4340Ph}
\maketitle

\section{Introduction}
Elastic waves at kHz and MHz frequencies are widely used to evaluate the mechanical 
properties of structures, material and tissues.
In plates and shells, the elastic wave equation admits specific propagation modes,
 denoted Lamb waves [\cite{viktorov1967, royerdieulesaint}], which are related to the traction 
free condition on both sides of the medium. Depending on the purpose of the experiment,
 different measurement configurations have been proposed. From the dispersion of Lamb waves, 
obtained from pitch-catch measures repeated for different ranges, one has access to the velocities of bulk waves 
and to the thickness of the plate ( see for instance \cite{gao2003}). Other pulse-echo (or impact-echo) techniques
have been proposed to assess the mechanical properties of the plate, based on the simple and multiple reflection 
of bulk waves within the plate [\cite{krautkramer1990}] or on the resonance of high order Lamb modes [\cite{clorennec2007}].
Some studies also concern the dynamic evaluation of fatigue and/or crack growth 
(see for instance \cite{ihn2004, ing1996}). In this last application, several impulse responses of the medium
are acquired at different dates and are eventually compared to each others to monitor the medium.

All these techniques require the use of controlled sources and receivers. In the following, they are referred to as "active" experiments. Another idea has undergone a large development after the seminal experiments of \cite{weaver2001} (see for instance the review of \cite{wapenaar2006}, \cite{weaver2006a} or \cite{larose2006e}). By cross-correlating the incoherent noise recorded by two passive sensors, Weaver and Lobkis demonstrated that one could reconstruct the impulse response of the medium as if a source was placed at one sensor. This noise correlation technique (also referred to as "passive imaging", or "seismic interferometry" [\cite{schuster2004}]) requires
the use of synchronized sensors, and has the advantage of eliminating any controlled source. 

In section II of our paper, we compare the dispersion curves of flexural waves (Lamb waves in the low frequency regime) in a plexiglass plate obtained by an active (pitch-catch) experiment to the ones obtained in a passive experiment. During the passive acquisition, we deployed  synchronised receivers only, and used a random noise source: a 1~mm thick high pressure air jet produced by a can of compressed air [\cite{mcbride1976}]. The experiment conducted in a plexiglass plate show at the same time: dispersive ($A_0$) and non dispersive ($S_0$) waves, reverberations, and absorption. Since the earth's crust has similar properties for seismic waves, we believe that a plexiglass plate at kHz frequencies is a good candidate to built small scale seismic analogous experiments. The question of reconstructing not only the phase, but also the amplitude of the wave is addressed at the end of part II. This is of interest in seismology, where the role of absorption and attenuation in the correlation has not yet been subject to experimental investigation. In section III, we analyze the rate of convergence of the correlations to the real impulse response. The role of the record length, the central frequency, and the distance between the receivers is investigated.

\section{Experimental setup and dispersion curves}
To study actively and passively the dispersion of flexural waves, we built a laboratory experiment using a $1.5 m\times 1.5 m$ large, 0.6~cm thick plexiglass plate. The plate is laid on an open steel frame that supports the edges of the plate but leaves free the upper and lower sides. The traction-free condition is therefore achieved on both horizontal sides. The resulting dispersion relation that connects the pulsation $\omega$ and the wave-vector $k$  reads:

\begin{equation}\label{dispersion}
\frac{\omega^4}{c_s^4}=4k^2q^2 \left(1-\frac{p \tan (ph/2+\gamma)}{q\tan(qh/2+\gamma)}\right)
\end{equation}
where $p^2=\frac{\omega^2}{c_p^2}-k^2$ and $q^2=\frac{\omega^2}{c_s^2}-k^2$; $c_s$ (resp. $c_p$) is the shear (resp. compressional) velocity, and $h$ is the thickness of the plate. The parameter $\gamma$ equals $0$ for symmetric ($S$) modes, and $\pi/2$ for anti-symmetric ($A$) modes. In the low frequency regime, only two modes are solutions: they are labeled $A_0$ and $S_0$.

For both acquisitions we used broad-band miniature (3~mm radius) accelerometers (ref.\# 4518 from Bruel\&Kjaer). They show a flat response in the 20~Hz-70~kHz frequency range. We fixed our accelerometers on our plate using a hot chemical glue (phenyl-salicylic acid) that solidifies with cooling (below 43$^\circ$C). 

\begin{figure}
	\centering
		\includegraphics[width=8cm]{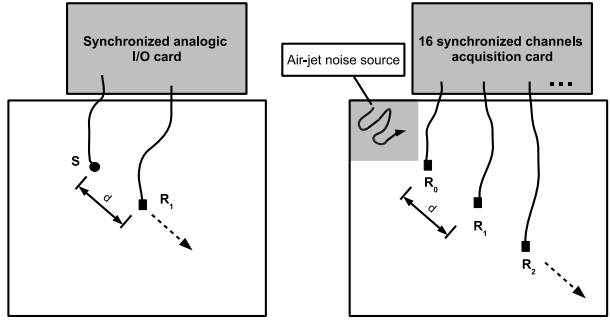}
		\label{fig1}
	\caption{Experimental setup. (a) In the active experiment (pitch-catch configuration), a broadband piezoelectric source $S$ emits a chirp that is sensed by a vertical accelerometer $R$ placed at a distance $d$. (b) In the passive experiment, a turbulent jet produced by a can of compressed air generates a white random noise recorded simultaneously at all sensors $R_i$. The jet is randomly sprayed over the area in gray.}
\end{figure}

\subsection{Active experiment}
In the active experiment, a source $S$ (a piezoelectric polymer) is in a corner of the plate, approximately 30~cm from each side and is emitting a 3~s chirp $s(t)$ in a linear range of frequencies $f$ from 1~kHz to 60~kHz. A receiver $R$ is initially placed at $d=$1~cm away from the source toward the center of the plate on a straight graduated line. After each acquisition, we moved the receiver a centimeter away from the source down the diagonal; we repeated this operation 100 times to cover 100~cm of the plate. The dynamic impulse response $h_d(t)$ of the plate is reconstructed for each distance $d$ by correlating the record $r_d(t)$ by the source chirp:

\begin{equation}
h_d(t)=r_d(t)\times s(t)
\end{equation}

\begin{figure}
	\centering
		\includegraphics[width=8cm]{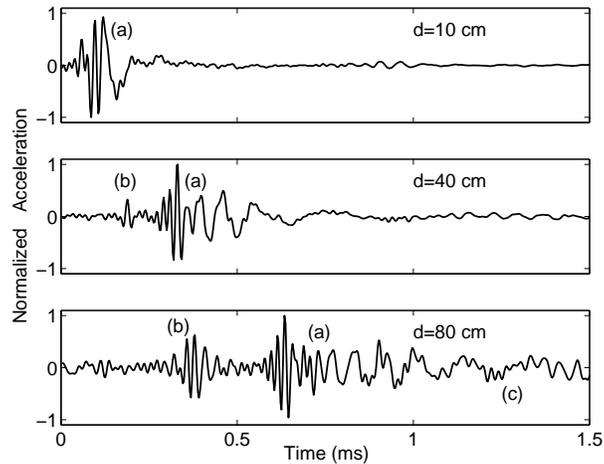}
		\caption{Source-receiver impulse response $h_d(t)$ for different distances $d$. (a) Dispersive anti-symmetric $A_0$ mode propagating at velocities approximately ranging from 150 to 900 m/s. (b) Non-dispersive $S_0$ mode propagating at $\approx$2400 m/s. (c) Reflexions from surrounding edges.}
	\label{fig2}
\end{figure}

As an example, impulse responses obtained for three different distances $d=10, 40, 80$~cm are displayed in Fig.~\ref{fig2}. The dispersive $A_0$ mode is dominating the record (a), but the non-dispersive $S_0$ wave (b) and reverberations from the edges of the plate (c) are also visible. The transit time in the plate is of the order of a few milliseconds. The absorption time of the plate that strongly depends on the frequency is of the order of a few hundreds of milliseconds. The signal-to noise in the experiment do not allow to record more than a few tens of millisecond. Records are therefore dominated by ballistic waves, but waves reverberated from the plate boundaries are also visible. \\

\begin{figure}
	\centering
		\includegraphics[width=8cm]{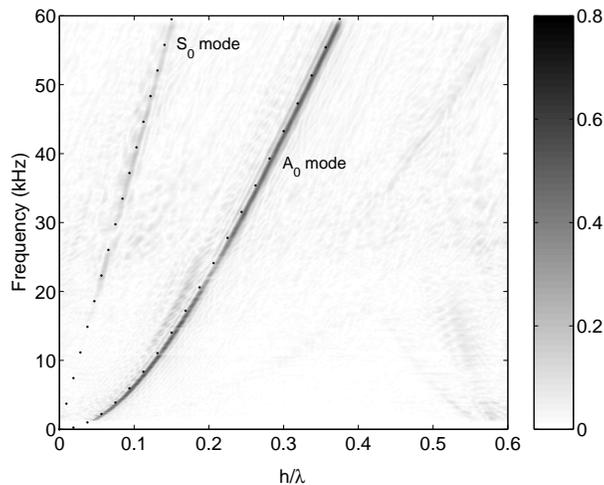}
		\caption{f-k transform of the source-receiver impulse responses. X-axis: dimensionless wavenumber ($h$ is the slab thickness). Dots are theoretical solutions of Eq.~\ref{dispersion}. }
	\label{fig3}
\end{figure}


 From the set of 100 impulse responses $h_d(t)$, a spatio-temporal Fourier (f-k) transform was applied. The resulting dispersion curves are displayed in Fig.~\ref{fig3}, showing the dispersive $A_0$ mode and the weaker non-dispersive $S_0$ mode. Theoretical dispersion curves are numerically obtained after Eq.~\ref{dispersion} (dots), they perfectly fit the experimental data for $c_p=3130$~m/s and $c_s=1310$~m/s. The dispersion curves are widely used to evaluate the mechanical properties of the plate, including thickness, presence of flaws... Nevertheless, the source-receiver configuration is sometimes at a disadvantage. As mentioned in the introduction, instead of using the conventional source-receiver configuration, it is possible to take advantage of random elastic noise to reconstruct the impulse response $h_d(t)$ between two sensors. This is performed in the following section. \\

\subsection{Passive experiment}\label{secB}
In the passive experiment, we removed the source and placed 16 receivers separated by 7 cm from each other on the graduated line. At the former position of the source, we placed the accelerometer $R_0$ that was kept fixed all through the experiment. As proposed by \cite{sabra2007}, we used the noise generated by a turbulent flow. As a noise source, we employed a dry air blower.  Note that, contrary to Sabra's work, our source was easy-to-handle. During the experiment, it was randomly moved to cover a 30~cm$\times$30~cm large area located between the corner of the plate and the former active source.  Note that the precise knowledge of the noise source is not necessary in the passive experiment. We also deployed an array of 16 receivers that allow for precise frequency-wave number analysis, and worked at higher frequencies, meaning a much thiner spatial resolution. We  focused our attention on the $A_0$ and $S_0$ direct arrivals.  Since we want to reconstruct direct waves, we chose to spray only at one end of the array of receivers (end-fire lobes described by \cite{roux2004} or coherent zones described by \cite{larose2006e}). Nevertheless, spraying elsewhere gives the same waveforms but requires much more data, thus much longer acquisitions (see section III). \\

The noise in the plate was created by spraying continuously for approximately T=10~s. The 16 receivers recorded synchronously this 10~s noise sequence, each record is labeled after its distance $d$ from $R_0$. Then receivers $R_1$ to $R_{15}$ were translated one centimeter down the graduated line and the 10~s acquisition performed again. This operation was repeated seven times to cover the 100~cm of the diagonal with a pitch of 1~cm. We ended with a set of 106 $r_{d=0..105}(t)$ records. The time domain cross-correlation between the receiver $R_0$ and the other receivers is processed afterward:

\begin{equation}
C_i(\tau)=\int_0^T r_0(t)r_i(t+\tau)dt.
\end{equation}

\begin{figure}
	\centering
		\includegraphics[width=8cm]{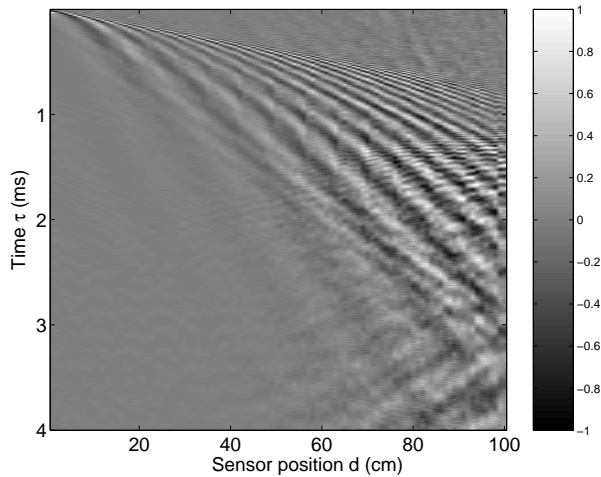}
	\caption{Passively reconstructed impulse responses (linear color bar, normalized amplitude).}\label{fig4}
\end{figure}

As mentioned in the introduction, this cross-correlation $C_d(\tau)$ is very similar to the impulse response $h_d(t)$ obtained in the pitch-catch experiment. Strictly speaking, the impulse response equals the time-derivative of the cross-correlation convolved by the source spectrum (which is almost flat here). Nevertheless, the time-derivative operation was not performed here since it does not change the spatio-temporal Fourier transform of the data. The series of 100 correlations is displayed in Fig.~\ref{fig4} as a time-distance plot. The dispersive $A_0$ mode is clearly visible, including reverberations at the edges. The symmetric $S_0$ mode is very weak and almost invisible in this figure. In the passive experiment, the size of the noise source is $\approx$1~mm large, which is quasi-punctual compared to the $A_0$ and $S_0$ wavelength. In that case, and taking into account the large difference in phase velocities, the energy ratio is clearly in favour of the $A_0$ mode, the $S_0$ mode is much weaker (and additionally less converged). In the active experiment the source was 6~mm large, thus exciting the $S_0$ mode more strongly and the $S_0$ mode was more visible in Fig.\ref{fig3}.
\\

\begin{figure}
	\centering
		\includegraphics[width=8cm]{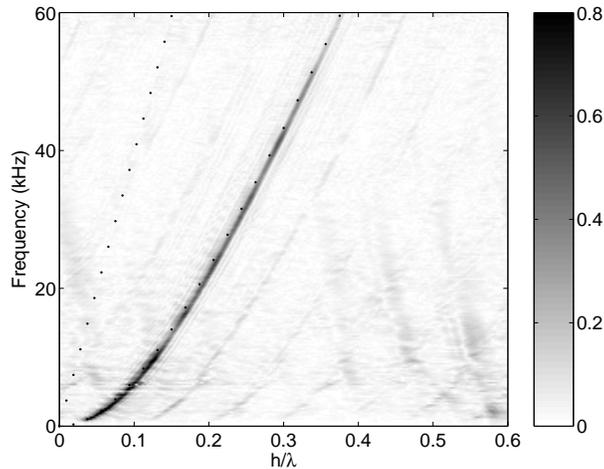}
	\caption{Spatio-temporal (f-k) Fourier transform applied to the passive data (linear color bar, normalized amplitude). The dispersion curve of the $A_0$ mode is perfectly matching the active experiment. $S_0$ mode is very weak, probably feebly excited by the point-like noise source. The dots show the theoretical dispersion curves as obtained in the active experiment.}\label{fig5}
\end{figure}

\begin{figure}
	\centering
		\includegraphics[width=8cm]{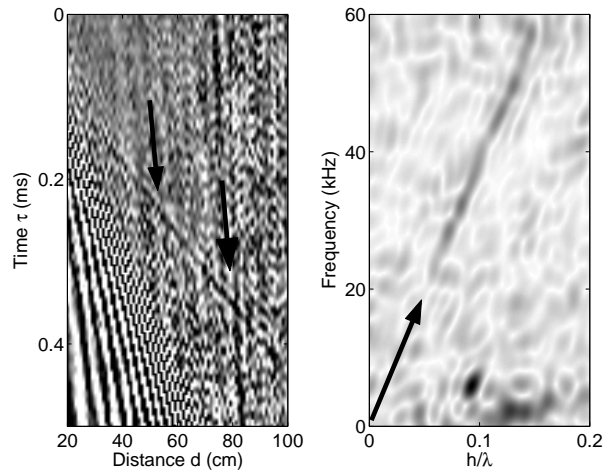}
	\caption{Left: zoom into the early part of the passive data (saturated color bar). The non-dispersive $S_0$ wave is pointed by the arrows. Right: f-k transform applied to the passive data after muting the $A_0$ mode arrivals (set to zero). The $S_0$ non-dispersive mode is now visible.}\label{fig6}
\end{figure}

Like in the active experiment, we computed the spatio-temporal Fourier transform of the set of 100 traces $C_d(\tau)$. The resulting dispersion curves are plotted in Fig.~\ref{fig5}. The $A_0$ mode is clearly reconstructed, the agreement between the active, passive, and theoretical dispersion curves is perfect. Ghosts are also visible in this figure, and are due to 1) small  errors in positioning the sensors 2) the presence of reflections from the edges. As to Fig.~\ref{fig4}, the $S_0$ mode is not visible. We therefore zoomed into the early part of the correlations (Fig.\ref{fig6}-left) and muted all the data but the weak non-dispersive arrival. We then computed again the f-k transform of the data, as plotted in Fig.~\ref{fig6}-right. The straight line that emerges from the noise is exactly the $S_0$ dispersion curve, thus demonstrating that the weak arrival in Fig.~\ref{fig6}-left is indeed the $S_0$ mode. It is important to emphasize that the weakness of the $S_0$ mode is not a limitation of our correlation technique, but is connected to the noise generation.

\subsection{Reconstruction of the amplitudes}

Since now, the reconstruction of the phase (the arrival time of the wave) of the Green function by correlation of noise has been widely studied. Feeble attention was paid to the information carried by the amplitude of the reconstructed GF. We can report the first experiments of \cite{weaver2001} who noted that both the phase and amplitude of the signals were passively reconstructed, and also \cite{larose2006c} who used this amplitude information to study weak localisation without a source. In Sec.~\ref{secB}, we have seen that the phase information in the passive experiment perfectly matches the active data, what about the amplitude of the wave?\\

\begin{figure}[htbp]
	\centering
		\includegraphics[width=8cm]{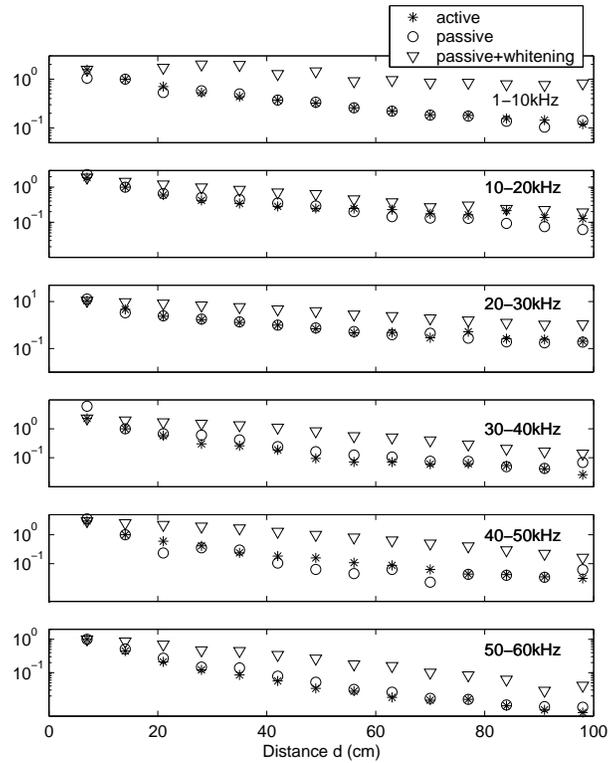}
	\caption{Amplitude decay obtained in an active experiment (stars) and by correlation (circles) without additional processing, and with whitening (triangles). Attenuation includes geometrical spreading and absorption.}
	\label{fig6bis}
\end{figure}

 To see if the amplitude decay is recovered using correlations, we plot on Fig.~\ref{fig6bis} the amplitude decay obtained with an active experiment (stars). In a homogeneous and lossless 2D plate, one would expect a decrease of the amplitude as $\propto 1/d^{0.5}$. Because of dispersion and the bandwidth used here (10 kHz), the exponent is slightly greater ($\propto 1/d^{0.6}$). To be more rigorous, absorption adds another exponential decay. From the active experiments, we found that the absorption time was about 260~ms at 60~kHz and 2000~ms at 1~kHz (similar to \cite{safaeinili1996}).\\

On Fig.~\ref{fig6bis}, we also plot the maximum of the amplitude of the correlations versus the distance $d$ between the sensors (circles). Active and passive amplitudes are comparable: the geometrical spreading of the wave is well recovered. Nevertheless, small discrepancies are visible and are due the imperfect or irregular coupling between the sensors and the plate. Estimating the absorption in the plate from the passive data is therefore more speculative with this experimental setup and will be subject to further investigations.\\

Nevertheless, in several applications [\cite{roux2004,shapiro2005,sabra2005b}], an operation is performed prior to the correlation in order to balance the contribution of all frequencies. The whitening operation, performed here in the 1-60~kHz frequency range, reads:
\begin{equation}
\tilde{r}_d(t)=IFFT \left(\frac{FFT\left(r_d(t)\right)}{\left|FFT\left(r_d(t)\right)\right|} \right),
\end{equation}

then the correlation $\tilde{r}_0(t)\times \tilde{r}_d(t)$ is evaluated. We compared the active amplitude with the amplitude of the correlations using whitened data (triangles in Fig.~\ref{fig6bis}). This time, though  a decay in the correlation is visible, we do not recover the decay of the actual GF. Note that 1-bit processing (correlate the sign of the records, see for instance \cite{larose2004b}) give similar disappointing results. Since whitening (or 1-bit) does not preserve the amplitude of the records, we stress that correlations will unlikely give the actual attenuation. This is of practical interest for applications like seismology. To summarize, in order to recover the amplitude by correlating the incoherent noise, we suggest that each record be filtered in a narrow frequency band, and then correlated without additional processing. The amplitude decay would therefore be estimated for each frequency, and attenuation derived. On the other hand, the whitening (or 1-bit) processing is valuable when the only phase information is targeted (tomography for instance).

\section{Rate of convergence of correlations}
Several authors [\cite{campillo2003a,derode2003a}] have experimentally noticed that the correlations $C_d(\tau)$ do not only contain the impulse response (the "signal") between the receivers but also show residual fluctuations that blur the weaker part of the signal. As soon as residual fluctuations are negligible compared to the reference impulse response, we consider that the correlations have converged. The convergence of correlations has been theoretically studied by \cite{weaver2005a} and \cite{sabra2005c}. We propose here to check the validity of their predictions with our experimental data. Of course, theoretical works apply to a perfectly diffuse wavefield. Here, the noise is generated over a delimited area. Nevertheless, as developed in section III.D, reverberations compensate the uneven distribution of source, and we believe that previous theories should apply to our experiment.

\subsection{Convergence with time $T$}
\begin{figure}
	\centering
		\includegraphics[width=8cm]{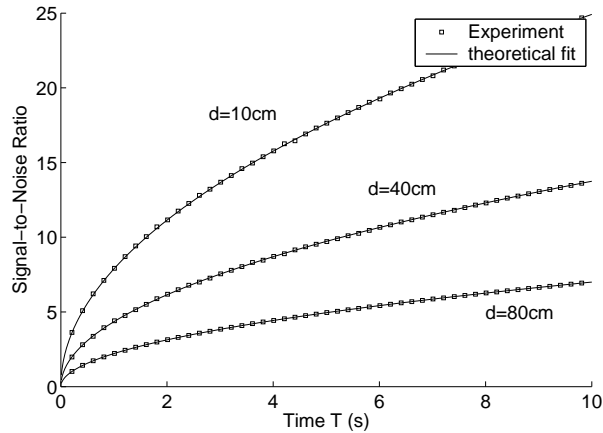}
	\caption{Signal-to-Noise Ratio of the correlation versus the duration T of the records. The theoretical fit for each distance $d$ is of the form $\alpha \sqrt{T}$ with $\alpha$ the fit parameter. }\label{fig7}
\end{figure}

To evaluate the degree of convergence of the correlations toward the real impulse response between the sensors, we split each noise record into N=1000 sub-records. Each sub-record lasts $\delta t=$10~ms. Correlations are processed for each sub-record and each couple of receivers $R_0-R_d$. We then compute the amplitude of the residual fluctuations:
\begin{equation}
\sigma_{d}(N,\tau)=\sqrt{\frac{\left\langle C_d(\tau)^2\right\rangle -\left\langle C_d(\tau)\right\rangle^2}{N-1}}
\end{equation}
where $\left\langle.\right\rangle$ represents an average over $N$ sub-records. The fluctuations were found to vary weakly with time $\tau$, we therefore average $\sigma_d(T=N\delta t,\tau)$ over $\tau$ to get a more robust estimation of the fluctuations $\sigma_d(T)$. To evaluate the "signal-to-noise ratio" (SNR) of the correlation, the maximum of each correlation is divided by the residual fluctuations $\sigma_d$. Experimental results are plotted in Fig.~\ref{fig7} for three distances $d$ and for the whole 1-60~kHz frequency range. As noticed theoretically and experimentally by several authors [\cite{weaver2005a,sabra2005c}], the SNR is found to increase like:

\begin{equation}
SNR= \alpha \sqrt{T}
\end{equation}
where $\alpha$ is the fit coefficient. This means that the longer the records, the better the reconstruction. The SNR increases as the square root of the amount of "information grain" contained in the records. This amount corresponds to the quantity $T\Delta f$ (where $\Delta f$ is the frequency bandwidth) and represents the number of uncorrelated pieces of information transported by the waves [\cite{derode1999,larose2004b}]. In our experiment, a few seconds was enough to get the direct arrival of the $A_0$ mode. \\

Another point visible in Fig.~\ref{fig7} is the decrease of the coefficient $\alpha$ with the distance $d$ between the pair of receivers. This means that the passive reconstruction of the impulse response is harder when the receivers are far apart. This was noticed in seismology for instance [\cite{shapiro2004,paul2005}]. A more precise estimation of the convergence with the distance $d$ is developed in the following subsection.

\subsection{Convergence with distance $d$}
\begin{figure}
	\centering
		\includegraphics[width=8cm]{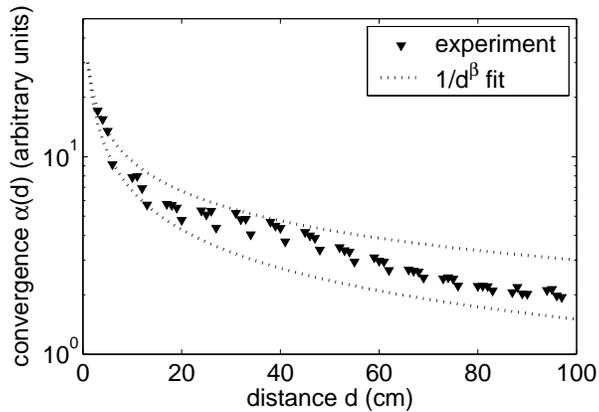}
	\caption{The fit coefficient $\alpha$ of the SNR is plotted versus the distance $d$ between the two correlated receivers. }\label{fig8}
\end{figure}

To study the dependence of the convergence rate on the distance $d$ between the passive sensors, each coefficient $\alpha$ was evaluated for each SNR curves (as in Fig.~\ref{fig7}) in the 1-60~kHz frequency range. The resulting $\alpha(d)$ is plotted in Fig.~\ref{fig8} and is delimited by two theoretical curves of the form:

\begin{equation} 
\alpha \propto 1/d^{\beta}.
\end{equation}
with $0.5<\beta<0.65$. At first sight, our result is similar to previous theoretical works that predicted a SNR evolution as $\sqrt{1/d}$ in a 2D space. Nevertheless, the theory was developed for a non-dispersive medium, where the wave spreading factor is $\sqrt{1/d}$.  In our dispersive plate, the exponent is slightly greater ($\approx 0.6$, see section II.A). The SNR decay seems to be driven by the decay of the actual Green function. note that the role of the absorption can not be evaluated in our experiment, and will be subject to further investigations.


\subsection{Convergence with frequency $f$}
\begin{figure}
	\centering
		\includegraphics[width=8cm]{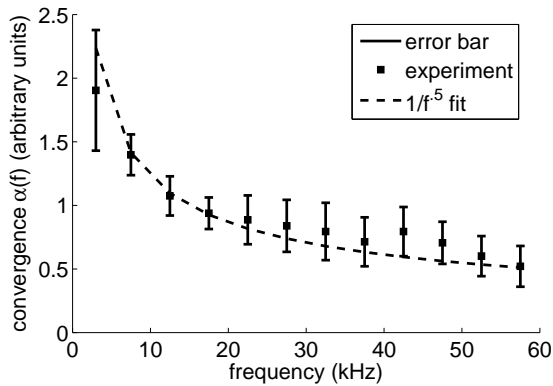}
	\caption{The fit coefficient $\alpha$ of the SNR is plotted versus the central frequency $f$ of the records. }\label{fig9}
\end{figure}

It is now commonly acknowledged that high frequencies are harder to reconstruct by cross correlation than low frequencies. We now propose to quantitatively evaluate the role of the central frequency $f$ of the record in the rate of convergence of the correlations. To that end, we performed several whitening operations in different frequency band (every 5~kHz, ranging from 1 to 60 kHz). Fluctuations where then evaluated for each frequency band and each sub-record. The experimental SNR was fitted by $\alpha(f) \sqrt{T}$, with $\alpha$ the fit parameter plotted in Fig.~\ref{fig9}. The prediction that $SNR\propto 1/\sqrt{f}$  is well recovered.

\subsection{Convergence with the noise sources position}
Several theoretical approaches have been proposed to model the reconstruction of the GF between passive sensors . Some invoke a perfect distribution of sources surrounding the medium to image [\cite{derode2003a,wapenaar2004}]. Others refer to a perfect diffuse wavefield, obtained either with scatterers or with thermal noise [\cite{weaver2001}]. Our experiment does not correspond to the first case, since the noise is generated in limited area. Does it correspond to the diffuse case? Because ballistic waves strongly dominate the records, this is not guaranteed and should be checked thoroughly! \\

\begin{figure}
	\centering
		\includegraphics[width=8cm]{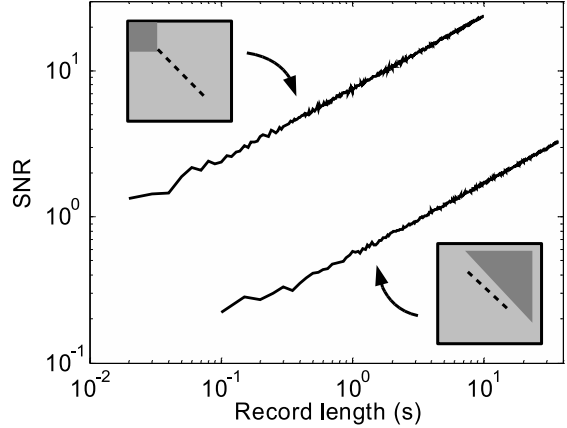}
	\caption{Rate of convergence of the correlation (SNR) with the record time T. The air-jet is randomly sprayed over the dark gray areas, doted line is the array of receivers. Upper left plot: sources located on the side of the array of receivers. Lower right plot: sources located in others places.}
	\label{fig11}
\end{figure}

To answer this point, we chose to spray over an area in the side of the array of receivers (see the dark gray area in Fig.~\ref{fig11}). If only ballistic waves were present, the direct waves along the receiver line would not be reconstructed. This was not the case here: after an integration over a 35~s long record, we could reconstruct the same waveforms as in he active experiment. The only difference is the rate of convergence of the correlations: it was found to be much slower (see SNR in Fig.~\ref{fig11} in log-log scale). This proves the elastic wavefield to be (at least slightly) diffuse in the plate employed in the experiment. To be more precise, the ballistic and the diffusive regime coexist in our plate. Right after a noise source occurs, the propagating wave paquet is in the ballistic regime. It turns into the diffusive regime after a dozen of reverberation. The first regime dominates the record; the second is weaker but not completely negligible. This last experiment is also a clear evidence that even with imperfectly diffusive wavefield, correlations of incoherent noise yield the Green function. This is of importance in applications like seismology where the medium is not always perfectly diffusive and the noise sources are not perfectly distributed. Figure~\ref{fig11} also demonstrates that when the reconstruction of the direct waves is expected, spraying over an area located in the direction of the array requires much less data than any other configuration.\\

\section{Conclusion and discussion}
In this article, we presented two experiments using Lamb waves detected by a series of accelerometers fixed on a plate. In the first experiment, a conventional source-receiver configuration was employed to construct the dispersion curves of the $A_0$ and $S_0$ flexural waves. In the second experiment, we only used receivers: we recorded the noise generated by a can of compressed air whose turbulent stream generates random excitations. Contrarily to an active experiment, the advantage here was that we neither needed to know the position of the noise source, nor to employ any electronics for the emission. By correlating the noise records, we reconstructed the impulse response between the sensors, and recovered the same dispersion curves as in the active experiment (except for the low frequency $S_0$ mode that was not excited by the thin turbulent jet).\\

 The amplitude decay of the actual $A_0$ mode with distance was successfully retrieved when correlating the raw records, but was not retrieved using either whitened or 1-bit records. This is of importance for applications where passive reconstruction of the attenuation of the media is envisioned. To summarize, whitening (or 1-bit) the data is helpful to reconstruct the phase of the GF (for imaging applications or f-k analysis), but should not be employed to reconstruct the amplitude.\\

Using our dense array of receivers, we also carefully studied the fluctuations of the correlations, which are connected to the rate of convergence of the correlations to the real impulse response. The Signal-to-Noise Ratio (or correlation-to-fluctuation ratio) obtained in our experiments is in agreement with previous studies [\cite{weaver2005a,sabra2005c}], both qualitatively and quantitatively. We showed that the reconstruction of the impulse response is better when: 1) we use long record (more data); 2) we employ close receivers; 3) we work at low frequency. The resulting SNR curves were best fitted by:
\begin{equation}
SNR= B \sqrt{\frac{T\Delta f c}{d^{1.1}f}}
\end{equation}
The parameter $B$, as defined by \cite{weaver2005a} was found to be $[0.25 \pm 0.02]$ (we take c=350m/s as a mean value for $A_0$ mode), which is much greater than seismic experiments [\cite{shapiro2004}]. This is in part due by the very uneven distribution of sources in our experiment, resulting in an anisotropic wave flux that accelerates the convergence of the correlation to the direct $A_0$ wave. This parameter B was found to be ten times smaller while spraying on one side of the array of receiver. Concerning the dependence with distance $d$, the decay matches the amplitude decay of the actual Green function. We emphasize that it is not a trivial term when dispersive waves are considered. A proper prediction for the SNR should account for the dispersion and the absorption.\\
 
 Practically speaking, in the 1-60~kHz frequency range, we obtained very good reconstruction of the impulse responses (SNR$\geq5$) for distances up to 100~cm with less than ten seconds of noise. This estimation is fast enough to be repeated continuously and provides a route for the passive monitoring of structures and materials. At seismic frequencies, correlations of ambient noise are already used to monitor active volcanoes [\cite{sabra2006, sensschonfelder2006, brenguier2007}]. As suggested by \cite{sabra2007} in a recent paper, we foresee similar applications in the field of on board passive structural health monitoring, in noisy environments such as aircrafts or ground vehicles, and also on civil engineering structures as bridges and buildings.

\section*{ACKNOWLEDGEMENTS}
We are thankful to D. Anache-Menier, P. Gouedard, Stefan Hiemer, L. Margerin, Adam Naylor, B. Van Tiggelen and R. L. Weaver for fruitful discussions and experimental help. Jean-Paul Masson is strongly acknowledged for technical help in designing the experiment. This work was funded by a french ANR "chaire d'excellence 2005" grant.


\end{document}